\documentclass[english,aps,prl,twocolumn,graphics,graphicx,superscriptaddress]{revtex4-1}
\usepackage[latin9]{inputenc}
\usepackage{amsmath,mathtools}
\usepackage{amssymb}
\usepackage{graphicx}
\usepackage{verbatim}
\usepackage{braket}
\usepackage{soul}
\usepackage{bm}
\usepackage{gensymb}
\usepackage{diagbox}
\usepackage{siunitx}
\usepackage[english]{babel}
\usepackage{mathtools}
\usepackage{newtxtext}  
\usepackage{amsmath}
\usepackage{amsfonts}
\usepackage{physics}
\usepackage[varbb,varg]{newtxmath} 
\usepackage{bm,dsfont,eucal}
\usepackage{graphicx}
\usepackage{pgf}

\begin{document}
\title{Halperin States of Particles and Holes in Ideal Time Reversal Invariant Pairs of Chern Bands and The Fractional Quantum Spin Hall Effect in Moir\'e MoTe$_2$}

\author{Inti Sodemann Villadiego}
\affiliation{Institut f\"ur Theoretische Physik, Universit\"at Leipzig, Br\"uderstra\ss e 16, 04103, Leipzig, Germany}

\date{\today} 

\begin{abstract}
An experiment in moir\'e MoTe$_2$ bilayers reported the first observation of a topologically ordered state with zero Hall conductivity and half of the edge conductance of a standard time-reversal invariant quantum spin Hall insulator. This state is believed to emerge at total filling one of a pair of bands with Chern numbers $C=\pm1$ related by time reversal symmetry. By viewing these bands as a pair of Landau levels with opposite magnetic fields, and starting from a parent magnet with one filled band, we demonstrate that a class of Halperin states constructed by adding particles to the empty Chern band and holes to the occupied Chern band have all the properties observed in MoTe$_2$. Remarkably, these states break time-reversal symmetry but have exactly zero Hall conductivity and helical edge conductance of $e^2/2h$. These states also feature a spinless composite fermion with the same charge as the electron but split equally between both valleys. In a standard Halperin 331 state, this particle would be a neutral Bogoliubov composite fermion. However, in our context this composite fermion is charged but remains itinerant because it is split into the two valleys that effectively experience opposite magnetic fields. The existence of such charged itinerant particles is a key difference between Landau levels with opposite magnetic fields and standard multi-components Landau levels, where all the itinerant particles are charge neutral, such as the magneto-roton of the Laughlin state or the Bogoliubov composite fermion of the Moore-Read state. When the electron density changes away from the ideal filling and these itinerant charged particles are added to the parent state, the disorder potential is less efficient at localizing them as compared to standard Lanadau levels. This can explain why the state in MoTe$_2$ did not display a robust Hall plateau upon changing the electron density.

\end{abstract}
\maketitle

\textit{\textcolor{blue}{Introduction}}. The dream to realize fractional quantum Hall states without magnetic fields has recently come true with the observations of fractional quantum anomalous Hall effect in twisted MoTe$_2$ bilayers (tMoTe$_2$) \cite{cai2023signatures,zeng2023thermodynamic,park2023observation,xu2023observation} and in pentalayer graphene on hBN \cite{lu2024fractional}. These striking discoveries follow a line of earlier observations in moir\'e twisted bilayer graphene of anomalous \cite{sharpe2019emergent,tseng2022anomalous}, anomalous integer-quantized \cite{serlin2020intrinsic,stepanov2021competing}, and fractional Chern insulators at relatively small fields \cite{xie2021fractional}, as well as integer-quantized anomalous Hall effect MoTe$_2$/WTe$_2$ bilayers \cite{li2021quantum,zhao2024realization} and in tMoTe$_2$ \cite{anderson2023programming}. These fractional anomalous quantum Hall states have been conceptualized as the result of spontaneous magnetism leading to particles polarizing into a flat Chern band where analogues of fractional quantum Hall states can become favorable \cite{ledwith2020fractional,repellin2020chern,abouelkomsan2020particle,wilhelm2021interplay,liu2021gate,li2021spontaneous,sheffer2021chiral,crepel2023anomalous,dong2023composite,goldman2023zero,morales2023magic,reddy2023toward,song2023phase,guo2023theory,dong2023theory,zhou2023fractional,dong2023anomalous,kwan2023moir,wang2024fractional,yu2024fractional,xu2024maximally}. 

\begin{figure}[h]
\centering
\includegraphics[width=0.40\textwidth]{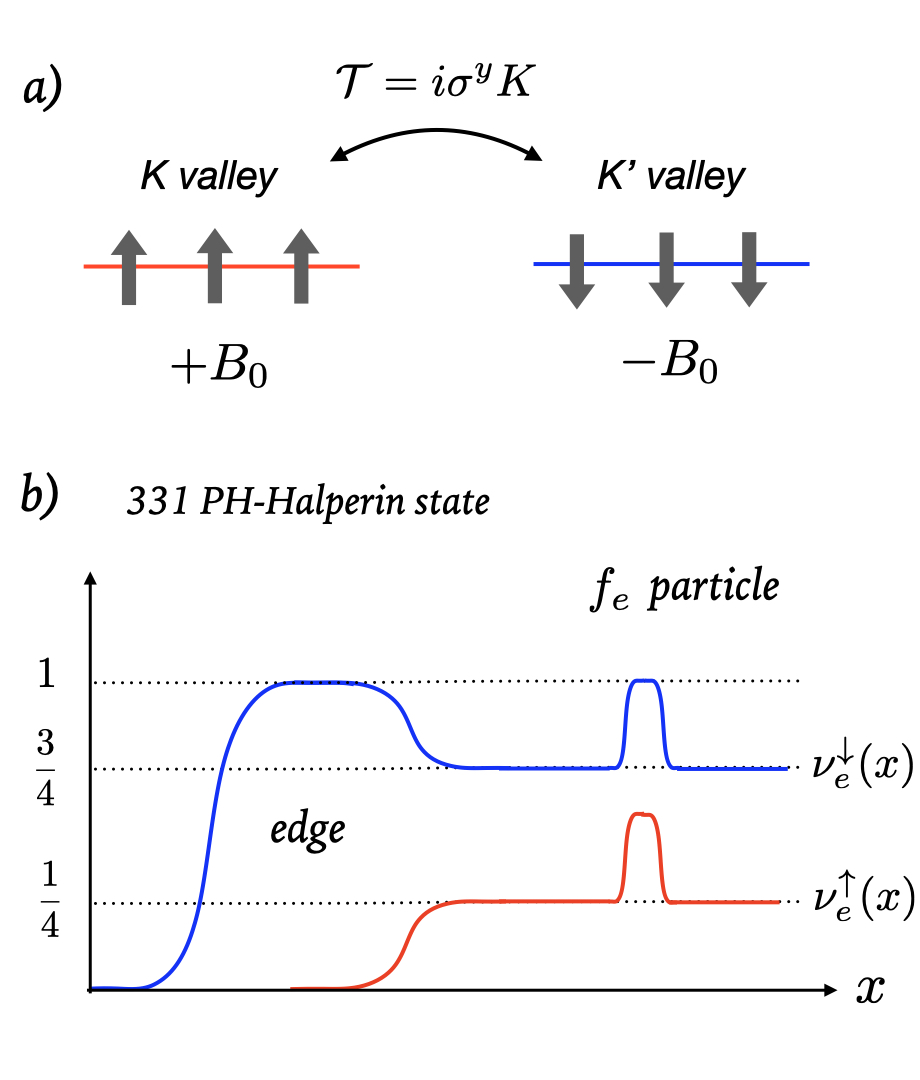}
\caption{(a) Two flat moir\'e bands from $K$ and $K'$ valleys and opposite spins and Chern numbers $C=\pm$ are simplified as a pair of Landau levels with opposite magnetic fields. (b) Profile of the PH-Halperin 331 state where holes of the $K'$ valley with spin $\downarrow$ and electrons of the $K$ valley with spin $\uparrow$ are added to a reference Integer Chern ferromagnet occupying valley $K'$. This state has a spinless itinerant composite fermion-like particle ($f_e$) with the same electric charge of the electron but split equally into each valley (see Fig.\ref{fig2}).}
\label{fig1}
\end{figure}

However, a recent remarkable experiment in tMoTe2 \cite{kang2024observation} has reported the first evidence of a fractionalized state that does not fit into the standard paradigm of a fractional quantum Hall states resulting from partially filling a Chern band. This experiment reported the striking observation of an insulating state with vanishing Hall conductivity, $\sigma_{xy}=0$, but with a quantized fractional edge conductance of $e^2/2h$, namely, the system behaves as if having an edge that is half of the standard time-reversal-invariant quantum spin Hall insulator  \cite{kane2005quantum,kane2005z,bernevig2006quantum}. The moir\'e bands of interest for this setting originate from the valence bands at $K$ and $K'$ valleys of MoTe$_2$ which are spin-split due to a large uniaxial spin-orbit field and related by time-reversal symmetry. Upon twisting two layers of MoTe$_2$ by a few degrees a  moir\'e pattern with a skyrmion texture of interlayer tunneling appears and gives rise to time-reveral pairs of flat bands with opposite Chern number originating from $K$ and $K'$ valleys \cite{WuPRL2019,devakul2021magic,wang2020correlated}. Related models also apply to bilayers of other transition metal dichalcogenides, such as WTe$_2$, where recently the double spin quantum Hall effect has been reported \cite{kang2024double}.

The presumptive fractional quantum spin Hall effect in tMoTe$_2$ is observed at a filling of $\nu=-3$ \cite{kang2024observation} (three holes per moire unit cell). At filling $\nu=-1$ the system behaves as an anomalous integer quantum Hall state with $\sigma_{xy}=e^2/h$ \cite{kang2024observation}, indicating that holes polarize onto a single moir\'e Chern band with unit Chern number. At filling $\nu=-2$ the system behaves as a time reversal invariant quantum anomalous spin Hall state with $\sigma_{xy}=0$ and conductance of $e^2/h$ per edge, indicating that holes fill equally a time-reversal invariant pair of bands with opposite Chern numbers $C=\pm 1$. At filling $\nu=-3$ the system displays an insulating state with vanishing Hall conductivity $\sigma_{xy}=0$ and edge conductance of $3e^2/2h$ per edge. We interpret this observation, as resulting from a non-trivial incompressible state constructed at half-filling of a pair of time-reversal invariant flat Chern bands residing on top of the simpler time-reversal-invariant vacuum of fully filled pair of Chern bands. Therefore, from here on we will focus on a single time-reversal invariant pair of flat Chern bands at an effective electron filling $\nu=1$ (see Fig.\ref{fig1}(a)). 

Reference \cite{kang2024observation} and a few subsequent theoretical proposals \cite{zhang2024vortex,shi2024excitonic,may2024theory,jian2024minimal}, interpreted the observation of vanishing Hall effect, $\sigma_{xy}=0$, as indicative that the ground state has time-reversal-invariant symmetry. We will, however, deviate from this point of view and propose instead a fully gapped incompressible state that spontaneously breaks time reversal symmetry but which has all the key properties observed in experiment, including the vanishing Hall conductivity and the fractional $e^2/2h$ helical conductance per edge. Our proposal is that tMoTe$_2$ realizes a correlated Halperin state \cite{halperin1983theory} with an equal number and electrons and holes added to a parent Ising Chern insulator at $\nu=1$, which we call a ``PH-Halperin" state. Our states are distinct from the Halperin states constructed in the standard quantum Hall setting \cite{halperin1983theory} because the electrons reside on Chern bands with opposite Chern number. They are also distinct from Halperin-like states of electrons discussed in the fractional quantum spin-Hall setting \cite{bernevig2006quantum} because they generically break time-reversal symmetry, and have unequal occupation of both valleys (see Fig.\ref{fig1}(b)).

Interestingly, we will see that PH-Halperin states with $\sigma_{xy}=0$ feature an emergent spin-less gapped composite-fermion-like particle that carries a total charge identical to the usual electron but equally split into halves in each spin-valley resolved Chern band (see Fig.\ref{fig1}(b)). Because this particle is split into occupying equally opposite Chern bands it behaves like the itinerant neutral dipolar particles of the standard quantum Hall setting (i.e. the magneto-roton of Laughlin states, or the Bogoliubov composite fermion of the Moore-Read state), even though it is charged. The existence of these itinerant charged particles is a unique feature of states constructed in pairs of Landau levels with opposite magnetic fields not present in ordinary Landau levels. These particles are harder to localize by the disorder potential as a result of an uncertainty relation between the average position and the relative distance of the two components residing in opposite Chern bands. The disorder potential tries to localize their average position, while interactions try to localize their relative distance. As a result there is a competition, and when interactions dominate, disorder will be less effective at pinning and localizing these itinerant charged particles and the corresponding analogue of the Hall plateau of these PH-Halperin states is expected to be more fragile relative standard quantum Hall states. This can explain why Ref.\cite{kang2024observation} did not observe a robust plateau, but instead a smooth deviation of the Hall conductivity away from zero as the electron density was changed away from the ideal filling where the incompressible state was observed.


\textit{\textcolor{blue}{Ideal model and symmetries}}. We consider a highly idealized model in which the Chern bands arising from both valleys of tMoTe$_2$ (labeled $\uparrow$ and $\downarrow$) are viewed as the $n=0$ Landau levels of a Hamiltonian for two species of particles experiencing opposite magnetic fields:
\begin{equation}\label{Hams}
H_\uparrow=\frac{(\mathbf{p}-\mathbf{A}_0- \delta \mathbf{A}^\uparrow)^2}{2m}, \ H_\downarrow=\frac{(\mathbf{p}+\mathbf{A}_0- \delta \mathbf{A}^\downarrow)^2}{2m},
\end{equation}


\noindent Here $\mathbf{A}_0$ is a spatially uniform magnetic field $\nabla \times \mathbf{A}_0=B_0 \mathbf{\hat{z}}$ of equal magnitude and opposite sign on the two valleys. The area of the moir\'e unit cell can be intepreted as the area of one flux quantum $a_{UC}=2\pi l_B^2= 2\pi/B_0$ \footnote{We use units of electron charge $q_e=1$ and $\hbar=1$, but will restore explicit units at the end in a few selected formulas.}. One crucial symmetry of the Hamiltonian and the states that we will consider is the $U_\uparrow(1)\times U_\downarrow(1)$ associated with separate particle number of each valley. This symmetry allows us to couple the system to two probe gauge fields, with vector potentials denoted by $\delta \mathbf{A}^{\uparrow,\downarrow}$, which we view as weak and slowly varying in space and time. Thus, the net magnetic and electric fields experienced by the particles in the two valleys can be different: 

\begin{align}\begin{aligned}
& \nabla \times (\mathbf{A}_0+\delta \mathbf{A}^{\uparrow})\equiv B_e^{\uparrow} \mathbf{\hat{z}}, \ {\bf E}_e^\uparrow \equiv -\partial_t \delta \mathbf{A}^{\uparrow},
\\
& \nabla \times (-\mathbf{A}_0+\delta \mathbf{A}^{\downarrow})\equiv B_e^{\downarrow} \mathbf{\hat{z}}, \ {\bf E}_e^\downarrow \equiv -\partial_t \delta \mathbf{A}^{\downarrow}.
\label{BEfields}
\end{aligned}\end{align}

\noindent Another important symmetry of our Hamiltonian is time reversal, $T=i\sigma^{y}K$ ($T^2=-1$), which exchanges the two valleys (present for $\delta \mathbf{A}^{\uparrow,\downarrow}=0$). However, the states of our interest will spontaneously break this symmetry. For concreteness it is useful to imagine that the system has an interacting Hamiltonian where particles in the same and in opposite valleys interact with different potentials $V_{\uparrow \uparrow}(\mathbf{r}_i-\mathbf{r}_j)$ and $V_{\uparrow \downarrow}(\mathbf{r}_i-\mathbf{r}_j)$  \cite{stefanidis2020excitonic}. We will imagine that the relevant physics emerges from projecting this Hamiltonian onto the $n=0$ pair of Landau levels from Eq.\eqref{Hams}, but we will not make explicit detailed use of the specific microscopic form of the interactions in our discussion.



\textit{\textcolor{blue}{Halperin states of particles and holes}}.
We start from an Ising Chern magnet as a reference vacuum in which particles fully occupy the valley with spin $\downarrow$ with corresponding Chern number $C=-1$. A common trick in this setting that allows to map the problem into a usual quantum Hall setting of two two-components with the same magnetic field, is to perform a particle-hole conjugation on the $\downarrow$ particles (see e.g. \cite{stefanidis2020excitonic,kwan2021exciton,kwan2022excitonic}). Thus, effectively, the $\uparrow$ particles and the $\downarrow$ holes behave like two species of particles in the same field $B_{e}^\uparrow =
 B_{h}^\downarrow=B_0$ in an $n=0$ Landau level with a total number of flux quanta $N_\phi=A/a_{UC}=B_0 A/2\pi$ (for $\delta \mathbf{A}^{\uparrow,\downarrow}=0$). This is a setting well studied in multi-layer and multi-component quantum Hall systems, but one important difference is that the sign of inter-flavor interaction, $V_{\uparrow \downarrow}$, would be flipped upon such partial particle-hole conjugation. 

In order to construct states with the same total electron density as the Ising Chern magnet reference vacuum, we add as many holes to the $C=-1$ band as electrons to the $C=1$ band, $N_e^{\uparrow}=N_h^{\downarrow}=N_\phi-N_e^{\downarrow}$. We will assume that these electrons and holes are forming a correlated Halperin $mmn$ state \cite{halperin1983theory} with wave-function:

\begin{equation} \label{PHHalp}
    \Psi^{PH}_{mmn} = \prod_{i<j}^{N_e^{\uparrow}} (z_i-z_j)^m \prod_{i<j}^{N_h^{\downarrow}} (w_i-w_j)^m \prod_{i,j}^{N_e^{\uparrow},N_h^{\downarrow}}  (z_i-w_j)^n ,
\end{equation}

\noindent where $m$ is an odd integer, $n$ is a non-negative integer, and we have omitted the standard exponential factors of $n=0$ Landau level wave-functions. Notice that for $n>0$ in the above states electrons of $\uparrow$ component have repulsive correlations with the holes of $\downarrow$ component, suggesting that in order to stabilize these states there needs to be an effective attraction between electrons of $\uparrow$ and $\downarrow$ components, or at least a reduction of the repulsion $V_{\uparrow \downarrow}$ relative to $V_{\uparrow \uparrow}$ so that electrons gain correlation energy relative to those they have in the Ising Chern magnet vacuum. Investigating the mechanism behind these energetics is an important problem, but in this study our goal will be instead to understand the properties of these PH Halperin states under the assumption that they are indeed the stable incompressible ground states of the system.

\textit{\textcolor{blue}{Bulk properties of PH-Halperin states}}. We begin by establishing a matrix generalization of the Streda relation between valley resolved magnetic fields and particle densities. To do so we hypothetically allow the net effective magnetic fields experienced by the $\uparrow$ electrons and $\downarrow$ holes, to be slightly different from the background field $B_0$ generated by the moir\'e potentials, by considering non-zero $\delta \mathbf{A}^{\uparrow,\downarrow}$ in Eqs.\eqref{BEfields}. The field experienced by the $\downarrow$ holes is $B_{h}^\downarrow=-B_{e}^\downarrow$. We also allow the number of particles to be different ($N_e^{\uparrow}\neq N_h^{\downarrow}$) relative to the ideal ground state. Let us imagine that the PH-Halperin state is placed on a manifold of area $A$ without boundaries such as the torus or the sphere. Matching the areas of the droplets of both spin components from Eq.\eqref{PHHalp} so that no fractional quasiparticles are added and the system remains in its deformed incompressible ground state under these sligthly modified conditions, leads to the following relations in the thermodynamic limit:

\begin{equation} \label{eq:PHHalp}
   \left(
\begin{array}{c}
 B_{e}^\uparrow \\
 B_{h}^\downarrow \\
\end{array}
\right)=2 \pi\left(
\begin{array}{cc}
 m & n \\
 n & m \\
\end{array}
\right) \left(
\begin{array}{c}
 n_{e}^\uparrow \\
 n_{h}^\downarrow \\
\end{array}
\right),
\end{equation}



\noindent with $n_{e}^\uparrow=N_e^{\uparrow}/A$, $n_{h}^\downarrow=N_h^{\downarrow}/A$. By converting back to electron variables, $B_{e}^\downarrow=-B_{h}^\downarrow$, $n_{e}^\downarrow=B_{h}^\downarrow/2\pi-n_{h}^\downarrow$, we obtain the following layer-resolved Streda relations: 

\begin{equation} \label{Stredaelectron}
   \left(
\begin{array}{c}
 n_{e}^\uparrow \\
 n_{e}^\downarrow \\
\end{array}
\right)=\frac{1}{2\pi}\left(
\begin{array}{cc}
 \frac{m}{m^2-n^2} & \frac{n}{m^2-n^2} \\
 \frac{n}{m^2-n^2} & \frac{m}{m^2-n^2}-1 \\
\end{array}
\right) \left(
\begin{array}{c}
 B_{e}^\uparrow \\
 B_{e}^\downarrow \\
\end{array}
\right).
\end{equation}


\noindent As we will see, a wealth of physical properties of the PH Halperin states can be derived from the above relations using a few reasonable physical assumptions. For example, the electron filling of the moir\'e bands for each spin component can be obtained by restoring the magnetic fields to the value generated by the moir\'e potential in the absence of perturbations ($B_{e}^\uparrow =
 -B_{e}^\downarrow=B_0$), and are given by:

\begin{equation} \label{fillings}
  \nu_e^{\uparrow }\equiv\frac{2 \pi n_e^{\uparrow}}{B_0} =\frac{1}{m+n}, \ \nu _e^{\downarrow}\equiv\frac{2 \pi n_e^{\downarrow}}{B_0}=1-\frac{1}{m+n}.
\end{equation}

\noindent Moreover the Eqs.\eqref{Stredaelectron} also govern the valley-resolved current density response to local electric fields in the bulk. This can be derived by imagining that the PH Halperin states are placed in a geometry without boundaries so that there is a full gap to all excitations and that the probe fields $\delta \mathbf{A}^{\uparrow,\downarrow}$ are varied weakly and slowly in time and space. Assuming that the ground state evolves adiabatically under such perturbations and that Eqs.\eqref{Stredaelectron} holds locally in space and time, one can derive a relation between local currents and electric fields. This follows from combining the continuity equations for the particle densities and recasting Faraday's laws as 2D continuity equations as follows:

\begin{equation} \label{continuity}
  \partial_t n_e^{\uparrow,\downarrow}=-\nabla \cdot {\bf j}_e^{\uparrow,\downarrow} , \
\partial_t B_e^{\uparrow,\downarrow}=-\nabla \cdot ({\bf E}_e^{\uparrow,\downarrow}\times \hat{{\bf z}}).
\end{equation}

\noindent By combining the above with Eqs.\eqref{Stredaelectron} and assuming that currents are locally orthogonal to electric fields so that the latter are dissipationless and do not perform work, one obtains the following valley resolved Hall conductivity matrix:

\begin{equation} \label{Hallmatrix}
   \left(
\begin{array}{c}
 {\bf j}_{e}^\uparrow \\
 {\bf j}_{e}^\downarrow \\
\end{array}
\right)=\frac{e^2}{h}\left(
\begin{array}{cc}
 \frac{m}{m^2-n^2} & \frac{n}{m^2-n^2} \\
 \frac{n}{m^2-n^2} & \frac{m}{m^2-n^2}-1 \\
\end{array}
\right) \left(
\begin{array}{c}
 {\bf E}_e^{\uparrow}\times \hat{{\bf z}} \\
 {\bf E}_e^{\downarrow}\times \hat{{\bf z}} \\
\end{array}
\right),
\end{equation}

\noindent where we have restored the explicit units of electrical conductivity for convenience. The above formula demonstrates that the PH Halperin states feature a non-trivial quantized Hall-drag response, whereby an electrical field driving only one valley can induce a quantized Hall current in the other valley, analogous to that of standard Halperin states in Landau levels \cite{yang1998hall}.  From the above we can get the valley-resolved electric currents in response to a net physical electrical field that acts identically on both valleys to be:

\begin{equation} \label{eq:PHHalp}
  {\bf j}_e^\uparrow=\frac{e^2}{h} \left(\frac{1}{m-n}\right) {\bf E} \times \hat{{\bf z}}, \
{\bf j}_e^\downarrow=\frac{e^2}{h} \left(\frac{1}{m-n}-1\right) {\bf E} \times \hat{{\bf z}}.
\end{equation}

\noindent and the net bulk electrical conductivity is therefore the sum of the above coefficients and given by:

\begin{equation}\label{sigmaxy}
  \sigma_{xy}=  \frac{e^2}{h} \left(\frac{2}{m-n}-1\right).
\end{equation}

 \noindent Interestingly, we see that the Hall conductivity of the PH-Halperin states is not simply the sum or the difference of the filling factors of the two valleys from Eq.\eqref{fillings}. And even more remarkably, the above implies that the subset of PH Halperin states satisfying $m=n+2$ have exactly zero Hall conductivity $\sigma_{xy}=0$, and  fractional-$1/2$ spin-resolved Hall conductivities of equal magnitude and opposite sign: 

\begin{equation} \label{updowncurrent}
  {\bf j}_e^\uparrow=-{\bf j}_e^\downarrow=\frac{e^2}{2h} {\bf E} \times \hat{{\bf z}}.
\end{equation}

\noindent In other words, their bulk electrical response would be exactly half of a standard time-reversal invariant quantum spin-Hall insulator state \cite{kane2005quantum,kane2005z,bernevig2006quantum} with valley number conservation in which the two valley resolved bands with opposite Chern would be fully occupied. However, notice that such PH Halperin states are not time-reversal invariant. This can be seen from Eq.\eqref{fillings} by noting that generically the fillings of the two components are different $\nu_e^{\uparrow}\neq \nu_e^{\downarrow}$, for example the PH 331 state the fillings are $\nu_e^{\uparrow}=1/4$ and $\nu_e^{\downarrow}=3/4$ (see Fig.\ref{fig1}(b)). 


\textit{\textcolor{blue}{Edge properties of PH Halperin states}}. Let us now analyze the properties of the PH-Halperin states at the edge. From analogy to fractional quantum Hall states of correlated holes, such as the $\nu=2/3$ particle-hole conjugate to the Laughlin state \cite{kane1994randomness,wen1991gapless,wen1990electrodynamical,wen1991edge,macdonald1990edge}, one expects a non-trivial edge profile in which the ideal bulk PH-Halperin state might be surrounded by a strip with fillings $(\nu_e^{\downarrow},\nu_e^{\uparrow})=(1,0)$, since the PH-Halperin droplet is carved out from a parent Ising-Chern magnet vacuum of the $\downarrow$ valley (see Fig.\ref{fig1}(a)), although more complex variants are certainly possible (see e.g.\cite{wang2013edge}). Given the complexities of the edge even in standard quantum Hall settings and considering the additional uncertainties about physical ingredients that are special to moir\'e materials, we will not attempt here to develop a detailed microscopic description of the edge. Instead we will follow the spirit of thermodynamic considerations such as those discussed in Ref.\cite{cooper1997thermoelectric}, and appeal to the assumption that in each edge there is good local thermodynamic equilibration between the particles in each of the valleys, but no inter-valley scattering processes that violate the conservation of the number particles in each valley, and thus we will assume that different chemical potentials for each valley at the edge are well defined.

\begin{figure}[h]
\centering
\includegraphics[width=0.45\textwidth]{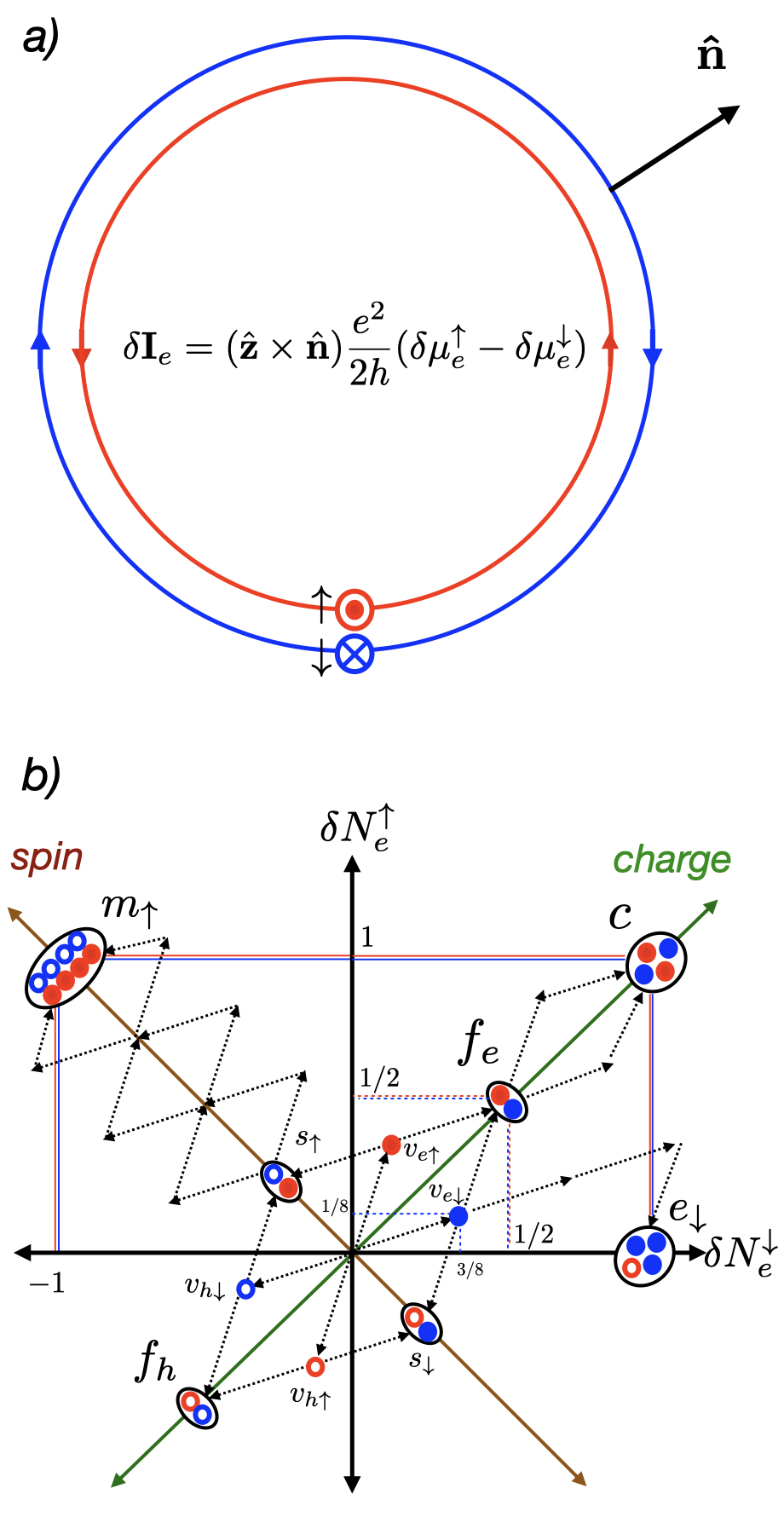}
\caption{(a) Depiction of the half-quantized helical curent transport at the edge of the PH-Halperin states with $m=n+2$. (b) Quasiparticle lattice of the 331 PH-Halperin state. The most elementary anyons are denoted by $v_{e\uparrow},v_{e\uparrow},v_{h\uparrow},v_{h\uparrow}$. Their bound states include charged spinless composite-fermion like particles $f_e$ and $f_h$, and the neutral spin-carrying semions $s_\uparrow$ and $s_\downarrow$. $c$ is the local inter-valley cooper-pair, $e_\downarrow$ is the local electron in valley $\downarrow$ and $m_\uparrow$ is a local charge-neutral spin-1 inter-valley magnon. Spinless particles  along the pure charge axis are itinerant charged particles, which is impossible in usual Landau levels where charged particles are also drifting particles which experience Lorentz force.}
\label{fig2}
\end{figure}

Let us begin by describing the edge when the system is globally in thermodynamic equilibrium (all edges included). There are two global chemical potentials $\mu_e^\uparrow,\mu_e^\downarrow$ for electrons in each valley. Particles in each valley are confined to some area $A$ by electrostatic potentials with an associated force ${\bf E}_0^{\uparrow,\downarrow}=-\hat{{\bf n}} E_0^{\uparrow,\downarrow}$, where $E_0^{\uparrow,\downarrow}>0$ and  $\hat{{\bf n}}$ is the local unit vector normal to the edge and pointing outwards from the sample (see Fig.\ref{fig2}). From Eq.\eqref{updowncurrent} we expect that there are net currents in equilibrium at the edge, which are flowing in opposite directions for the two spin components, and which for our conventions would be counterclockwise for the $\uparrow$ particles and clockwise for the $\downarrow$ particles (see Fig.\ref{fig2}). Following standard principles of equilibrium thermodynamics, we can define a grand-canonical free energy, $G=E-\mu_e^\uparrow N_e^{\uparrow}-\mu_e^\downarrow N_e^{\downarrow}$, where $E$ is the total energy of the system, whose differential is:

\begin{equation} \label{firstlaw}
  dG=TdS- N_e^{\uparrow} d\mu_e^\uparrow-N_e^{\downarrow} d \mu_e^\downarrow-A M_e^{\uparrow} d B_e^{\uparrow}-A M_e^{\downarrow} dB_e^{\downarrow} +\cdots  
\end{equation}

\noindent where $M_e^{\uparrow,\downarrow}$ are the particle number magnetization densities for the two valleys, and the $\cdots$ include the work differentials of variations of the Hamiltonian with respect to parameters other than $B_e^{\uparrow,\downarrow}$. From the above we obtain the following Maxwell relations: 

\begin{equation} \label{maxwell}
 \frac{\partial M_e^{j}}{\partial \mu_e^{i}}  = \frac{\partial n_e^{i}}{\partial B_e^{j}} , \ i,j \in \{\uparrow,\downarrow\}.
\end{equation}

\noindent Now from the assumption of incompressibility of the bulk, one concludes that any changes of the magnetizations, $\delta M_e^{\uparrow,\downarrow}$ in response to small variations of the chemical potentials $\delta \mu_e^{\uparrow,\downarrow}$ which remain within the bulk gap, must arise entirely from variations of the currents localized at the edges of the system. By integrating the relation between magnetization and current densities ${\bf j}_e^{\uparrow,\downarrow}=  \nabla \times (\hat{{\bf z}} M_e^{\uparrow,\downarrow}) $, one gets the relation between edge currents and magnetization ${\bf I}_e^{\uparrow,\downarrow}= M_e^{\uparrow,\downarrow} \hat{{\bf z}}\times \hat{{\bf n}} $ \cite{cooper1997thermoelectric}. Then by computing the right-hand-side of Eq.\eqref{maxwell} from Eq.\eqref{Stredaelectron}, one obtains the relation between variations of edge currents and global chemical potentials:

\begin{equation} \label{Stredaedge}
   \left(
\begin{array}{c}
\delta {\bf I}_{e}^\uparrow \\
\delta {\bf I}_{e}^\downarrow \\
\end{array}
\right)= (\hat{{\bf z}}\times \hat{{\bf n}}) \frac{e^2}{h} \left(
\begin{array}{cc}
 \frac{m}{m^2-n^2} & \frac{n}{m^2-n^2} \\
 \frac{n}{m^2-n^2} & \frac{m}{m^2-n^2}-1 \\
\end{array}
\right) \left(
\begin{array}{c}
 \delta \mu_e^{\uparrow} \\
 \delta \mu_e^{\downarrow} \\
\end{array}
\right),
\end{equation}

\noindent Now, following the spirit of Landauer-B\"uttiker approach \cite{landauer1970electrical,buttiker1986four}, in a transport experiment one expects that different edges connecting various voltage leads generally have different chemical potentials for the two spins components. Since the net direction of the $\uparrow,\downarrow$ edge currents is opposite, we assume that the chemical potential of these modes is in equilibrium with the voltage lead from which each of them emanates, in analogy to the standard Landauer-B\"uttiker picture for the quantum-spin-Hall effect \cite{kane2005quantum}. In particular we see from the above that the total electric current at the edge, $\delta {\bf I}_{e}=\delta {\bf I}_{e}^\uparrow+\delta {\bf I}_{e}^\downarrow$, is:

\begin{equation}\label{edgeI}
  \delta{\bf I}_{e}=(\hat{{\bf z}}\times \hat{{\bf n}}) \frac{e^2}{h} \left(\frac{1}{m-n}\right) \delta \mu_e^{\uparrow}-(\hat{{\bf z}}\times \hat{{\bf n}}) \frac{e^2}{h}\left(1-\frac{1}{m-n}\right) \delta \mu_e^{\downarrow},
\end{equation}

\noindent and, therefore, we see that for the special PH Halperin states of our interest with $m=n+2$ each edge behaves as a helical edge conductor with opposite current directions for each spin and fractional conductance $e^2/2h$, namely half of the usual quantum spin Hall effect, in agreement with experiment \cite{kang2024observation}: 

\begin{equation}\label{edgeI}
  \delta{\bf I}_{e}=(\hat{{\bf z}}\times \hat{{\bf n}}) \frac{e^2}{2h} (\delta \mu_e^{\uparrow}-\delta \mu_e^{\downarrow}).
\end{equation}

\textit{\textcolor{blue}{Bulk quasiparticles of PH Halperin states}}. There is a key distinction between standard multicomponent systems in Landau levels with the same magnetic field, and our current setting of pairs of Landau levels with opposite magnetic fields which becomes evident when we analyze their quasi-particles. In standard Landau levels only fully neutral particles are itinerant and can move in straight trajectories in the presence of magnetic field. Such itinerant neutral particles include the excitons in integer quantum Hall ferromagnets, the magneto-roton in the Laughlin state, and the Bogoliubov composite fermion in paired states such as the Moore-Read or the standard 331 Halperin state. In the current setting, however, it is possible to have charged quasi-particles that are itinerant, when the charge of the particle is equally split between the two valleys that experience opposite magnetic fields. The converse of this is also possible, namely in the current setting there are neutral particles that behave as the ordinary charged particles in standard Landau levels, as demonstrated for the inter-valley excitons of the Ising Chern magnet which have exactly opposite charges in the two Landau levels with opposite magnetic field, as discussed in Refs.\cite{stefanidis2020excitonic,kwan2021exciton,kwan2022excitonic}. Therefore, we would like to introduce a notion to distinguish particles not only as charged and neutral, but also as {\it itinerant} or {\it drifting} quasiparticles. Quasi-particles carry definite valley charges, which we denote by a vector $(\delta N_e^{\uparrow},\delta N_e^{\downarrow})$ indicating their total electric charge in each of the two valleys (in units where the electron charge is 1). Thus the total quasi-particle electric charge is $\delta N_e^{\uparrow}+\delta N_e^{\downarrow}$. Now, in the context of a pair of Landau levels where valleys have opposite magnetic fields, we will say that  a quasi-particle is itinerant if their valley polarization is zero, namely if:

\begin{equation}\label{edgeI}
  \delta N_e^{\uparrow}  =\delta N_e^{\downarrow}, \ {\rm (itinerant \ quasiparticle),}
\end{equation}

\noindent otherwise we will call the quasiparticle a drifting quasiparticle. For example, when we add quasiparticles to the trivial vacuum of fully empty valleys, the electrons would have numbers $(\delta N_e^{\uparrow},\delta N_e^{\downarrow})=(1,0)$ or $(\delta N_e^{\uparrow},\delta N_e^{\downarrow})=(0,1)$, and thus they would be charged drifting particles, whereas a Cooper-pair-like bound state of two electrons with  $(\delta N_e^{\uparrow},\delta N_e^{\downarrow})=(1,1)$ would be a charged itinerant particle. Conversely, an ordinary exciton with $(\delta N_e^{\uparrow},\delta N_e^{\downarrow})=(1,-1)$, which can be added to the Ising Chern magnet vacuum, would be a neutral drifting particle.

Since the PH Halperin states are abelian topologically ordered states, their topological properties can be captured by an abelian Chern-Simons theory with a K-matrix and layer resolved charge vectors \cite{wen2004quantum} given by: 

\begin{equation} \label{Kmatrix}
 K= \left(
\begin{array}{ccc}
 m & n & 0 \\
 n & m & 0 \\
 0 & 0 & -1 \\
\end{array}
\right), \ q_\uparrow=\left(
\begin{array}{c}
 1 \\
 0 \\
 0 \\
\end{array}
\right),\ q_\downarrow=\left(
\begin{array}{c}
 0 \\
 -1 \\
 1 \\
\end{array}
\right).
\end{equation}

\noindent All the bulk quasiparticles are expected to be fully gapped and can be labeled by vectors of integers ${\it{\bf l}}^T=(l_1,l_2,l_3)$. Their self-exchange statistical angle (topological spin) and layer resolved charges are given by $\theta_{\bf l}=\pi {\bf l}^T K^{-1} {\bf l}$, $\delta N_e^{\uparrow/\downarrow}=q_{\uparrow/\downarrow}^T K^{-1} {\bf l}$, while their braiding statistics (statistical phase of a particle after full loop around another particle) is $\theta_{{\bf l},{\bf l'}}=2 \pi {\bf l}^T K^{-1} {\bf l'}$ \cite{wen2004quantum}. The most elementary anyons are:

\begin{equation} \label{eq:PHHalp}
 v_{e\uparrow}=\left(
\begin{array}{c}
 1 \\
 0 \\
 0 \\
\end{array}
\right),v_{h\uparrow}=\left(
\begin{array}{c}
 -1 \\
 0 \\
 0 \\
\end{array}
\right), v_{e\downarrow}=\left(
\begin{array}{c}
 0 \\
 -1 \\
 0 \\
\end{array}
\right), v_{h\downarrow}=\left(
\begin{array}{c}
 0 \\
 1 \\
 0 \\
\end{array}
\right).
\end{equation}

\noindent where $v_{e\uparrow}$,$v_{h\uparrow}$ denote the Laughlin-type quasi-electron and quasi-hole particles in the $\uparrow$ component, and similarly, $v_{e\downarrow}$,$v_{h\downarrow}$ the corresponding quasi-particles for $\downarrow$ component. All of these quasi-particles are anyons with fractional self-exchange statistical angle $\theta_v=\pi m/(m^2-n^2)$. The quasi-electrons have a bump of electron density relative to the background PH Halperin ground state, and the quasi-holes a depletion, so that the layer-resolved charges of these quasi-particles are:

\begin{equation} \label{vecharges}
 v_{e\uparrow}: (\delta N_e^{\uparrow},\delta N_e^{\downarrow})= \left(\frac{m}{m^2-n^2},\frac{n}{m^2-n^2}\right).
\end{equation}

\noindent The corresponding layer resolved charges of the quasi-electron $v_{e\downarrow}$ are obtained by swapping $\uparrow \leftrightarrow \downarrow$ in the above formula, and those of the quasi-holes $v_{h\uparrow},v_{h\downarrow}$ are minus those of the corresponding quasi-electrons. From the above, we see that the $v$-type anyons are drifting quasi-particles because they have un-equal occupation of the two valleys whenever $m\neq n$, such as in the PH $331$ state. Interestingly, we see that the total charge of these quasi-particles is:

\begin{equation} \label{eq:PHHalp}
 v_{e\uparrow}: \ \delta N_e= \frac{1}{m-n},
\end{equation}

\noindent And therefore, for the PH-Halpering states with $m=n+2$, these quasi-particles carry 1/2 of the electron charge which is unequally split between both valleys according to Eq.\eqref{vecharges}. This contrasts with the standard Halperin states in multicomponent Landau levels (LLs), whose smallest quasiparticles have charge $(\delta N_e)_{\rm LLs}= 1/(m+n)$ which is split as $(\delta N_e^{\uparrow},\delta N_e^{\downarrow})_{\rm LLs}= (m/(m^2-n^2),-n/(m^2-n^2)))$. For example the standard, 331 Halperin state has smallest quasiparticles with total charge $1/4$ which is split as $(3/8,-1/8)$, whereas our 331 PH Halperin state has a smallest quasiparticles with total charge $1/2$ which is split as $(3/8,1/8)$. This difference can be understood as a result of the particle-hole conjugation of the $\downarrow$ component, which basically maps the layer pseudospin polarization of the quasiparticle in the usual Landau levels onto the total quasiparticle charge in the current setting: $(\delta N_e^{\uparrow}-\delta N_e^{\downarrow})_{\rm LLs}\rightarrow (\delta N_e^{\uparrow}+\delta N_e^{\downarrow})$. 

\noindent Let us now consider some of the quasi-particles that are obtained as bound states of pairs of the above elementary anyons. The ${\bf l}$ vectors of these bound states are obtained by adding those of the elementary anyons, and we label them as:

\begin{equation} \label{eq:PHHalp}
 f_{e}=v_{e\uparrow} \times \ v_{e\downarrow}  =\left(
\begin{array}{c}
 1 \\
 -1 \\
 0 \\
\end{array}
\right), \ f_{h}=v_{h\uparrow} \times \ v_{h\downarrow}  =\left(
\begin{array}{c}
 -1 \\
 1 \\
 0 \\
\end{array}
\right), 
\end{equation}

\begin{equation} \label{sanyons}
 s_{\uparrow}=v_{e\uparrow} \times \ v_{h\downarrow}  =\left(
\begin{array}{c}
 1 \\
 1 \\
 0 \\
\end{array}
\right), \ s_{\downarrow}=v_{h\uparrow} \times \ v_{e\downarrow} =\left(
\begin{array}{c}
 -1 \\
 -1 \\
 0 \\
\end{array}
\right).
\end{equation}

\noindent The layer-resolved charges and the self-statistical exchange phase of these quasi-particles are:

\begin{equation} \label{eq:PHHalp}
 f_{e}: (\delta N_e^{\uparrow},\delta N_e^{\downarrow})= \left(\frac{1}{m-n},\frac{1}{m-n}\right), \ \theta_f= \frac{2 \pi }{m-n},
\end{equation}

\begin{equation} \label{eq:PHHalp}
 s_{\uparrow}: (\delta N_e^{\uparrow},\delta N_e^{\downarrow})= \left(\frac{1}{m+n},-\frac{1}{m+n}\right), \ \theta_s= \frac{2 \pi }{m+n}.
\end{equation}

\noindent The particle $f_{h}$ has the same statistics and opposite layer-resolved charges of $f_{e}$, while $s_{\downarrow}$ has the same statistics and opposite charges of $s_{\uparrow}$. Therefore, we see that the $f_e,f_h$ are spinless particles with equal occupations of both layers and therefore are also charged itinerant particles, while the $s_{\uparrow},s_{\downarrow}$ are neutral spinful particles with opposite occupations of layers, and thus are drifting particles.

\noindent Remarkably, for the special class of PH-Halperin with $m=n+2$ that are consistent with the experiments of Ref.\cite{kang2024observation}, the $f_e,f_h$ particles are spinless fermions ($\theta_f= \pi $) with the same charge of an ordinary electron and hole respectively, but fractionalized onto halves residing on each spin component $(\delta N_e^{\uparrow},\delta N_e^{\downarrow})= \pm (1/2,1/2)$. Despite being charged these particles are therefore intinerant quasiparticles. Moreover their fermionic statistics imply that if a finite density of these particles is added to the PH-Halperin state vacuum, they can naturally form a compressible metallic itinerant state. In contrast for these special states with $m=n+2$, the $s_{\uparrow},s_{\downarrow}$ have in general fractional self-statistics ($\theta_s= \pi/(n+1) $) and are drifting non-intinerant particles.

The spinless fermions $f_e,f_h$ are non-local emergent particles and should not be confused with the microscopic electron or hole, even though they have the same statistics and total charge. This can be seen for example by noting that the statistical phase for moving $f_e$ around a loop encircling the anyon $v_{e\uparrow}$ is $\theta_{f_e,v_{e\uparrow}}=2 \pi/(m-n)$. For the special states with $m=n+2$ this is $\theta_{f_e,v_{e\uparrow}}=\pi$, which implies that this fermion $f_e$ sees the anyon $v_{e\uparrow}$ as an effective $\pi$-flux Abrikosov-like vortex. Such resemblance to the quasi-particles of a superconductor is not a coincidence; it is known that in ordinary two-component Landau levels the Halperin states with $m=n+2$ can be thought of as an inter-layer paired state of composite fermions \cite{read2000paired,haldane1988spin,greiter1992paired,ho1995broken,dimov2008spin,sharma2024composite,read1996quasiholes}. In ordinary Landau levels the $f$ particles would have been the charge neutral Bogoliubov fermions descending from the composite fermion upon pairing, and would have been an itinerant particle with opposite charges in the two layers. However, as we discussed before, our setting of valleys with opposite magnetic fields maps into the traditional Landau level setting after particle-hole conjugating one of the valleys. Under such transformation, the Bogoliubov composite fermion becomes a charged particle but retains its itinerant character. 

To close this section we would like to connect the above fractionalized quasi-particles to the standard local microscopic particles. The standard electrons and holes with valley-spins $\uparrow$ and $\downarrow$ can be obtained as:

\begin{equation} \label{electrons}
 e_{\uparrow}=(v_{e\uparrow})^m \times \ (v_{h\downarrow})^n  =\left(
\begin{array}{c}
 m \\
 n \\
 0 \\
\end{array}
\right), \ e_{\downarrow}=(v_{h\uparrow})^n \times \ (v_{e\downarrow})^m =\left(
\begin{array}{c}
 -n \\
 -m \\
 0 \\
\end{array}
\right),
\end{equation}

\begin{equation} \label{electrons}
 h_{\uparrow}=(v_{h\uparrow})^m \times \ (v_{e\downarrow})^m =\left(
\begin{array}{c}
 -m \\
 -n \\
 0 \\
\end{array}
\right), \ h_{\downarrow}=(v_{e\uparrow})^n \times \ (v_{h\downarrow})^m  =\left(
\begin{array}{c}
 n \\
 m \\
 0 \\
\end{array}
\right),
\end{equation}

\noindent The above are local fermions, with charges $(\delta N_e^{\uparrow},\delta N_e^{\downarrow})$ respectively given by $\{(1,0),(0,1),(-1,0),(0,-1)\}$. Interestingly, the inter-valley Cooper-pair-like boson obtained as the bound state of two ordinary electrons in the two valleys can be alternatively also obtained as the bound state of an $(m-n)$ multiple of $f_e$ composite-fermion-like particles:

\begin{equation} \label{cooper}
 c=e_{\uparrow} \times \ e_{\downarrow}  = (f_e)^{m-n} =\left(
\begin{array}{c}
 m-n \\
 n-m \\
 0 \\
\end{array}
\right),
\end{equation}

\noindent The above can be recognized as such inter-valley Cooper-pair because it is a self-boson which is local (i.e. has trivial full braiding with any other quasi-particle modulo $2\pi$) and carries charges $(\delta N_e^{\uparrow},\delta N_e^{\downarrow})=(1,1)$. For the PH-Halperin states with $m=n+2$ that are consistent with experiments, Eq.\eqref{cooper} can be interpreted as saying that a bound state of a pair $f_e$ composite-fermions forms an ordinary Cooper-pair, or alternatively, that the $f_e$ composite-fermion is one half of an ordinary Cooper-pair. We see that the Cooper pair is therefore the simplest fully local itinerant charged particle in these states.

\noindent Conversely, the inter-valley magnon-exciton can be obtained either as a bound state of an $\uparrow$ electron and a $\downarrow$ hole or alternatively as an $(m+n)$ multiplet of the $s$ anyons from Eq.\eqref{sanyons}, as follows:

\begin{equation} \label{cooper}
 m_{\uparrow}=e_{\uparrow} \times \ h_{\downarrow}  = (s_\uparrow)^{m+n} =\left(
\begin{array}{c}
 m+n \\
 m+n \\
 0 \\
\end{array}
\right),
\end{equation}

\noindent and, analogously, we also have a valley-spin raising particle:

\begin{equation} \label{cooper}
 m_{\downarrow}=e_{\uparrow}  \times \ h_{\downarrow}   = (s_\downarrow)^{m+n} =\left(
\begin{array}{c}
 -m-n \\
 -m-n \\
 0 \\
\end{array}
\right).
\end{equation}

\noindent these local magnon-like particles carry charges $(\delta N_e^{\uparrow},\delta N_e^{\downarrow})$ given respectively by $(1,-1)$ and $(-1,1)$, and thus are neutral spin-1 bosons, but are drifting (not intinerant) particles. We see, therefore, that the PH-Halperin states realize a beautiful and colorful pattern of spin-charge separation.

\textit{\textcolor{blue}{Localization of itinerant vs drifting particles and the robustness of the Hall plateau}}. We believe that the existence of itinerant charged particles can have important consequences on the robustness of the Hall plateau when the system is doped away from the precise filling of the ideal incompressible state. To argue for this, we begin by reviewing the mechanism behind the existence of a Hall plateau for a standard Laughlin or Halperin states in a usual Landau levels in the same magnetic field. The ideal fractional quantum Hall fluids exist at some precise fractional proportion between the electron density and the magnetic field. However, when the magnetic field or electron density are slightly changed away from this exact proportion, an experimental Hallmark of the fractional quantum Hall effect is that $\sigma_{xy}$ remains precisely quantized at the value associated with the ideal fluid as if this proportion had not changed. This occurs because the excess particles away from the ideal filling are accommodated in the form of Laughin-like quasi-particles added on top of the ideal vacuum. In standard Landau levels, these quasi-particles are charged and therefore drifting (non-itinerant) and thus easily localized and pinned by the disorder potential. For sufficiently small densities of these quasi-particles they remain pinned at disconnected locations and surrounded by the ideal Laughlin-like fluid with its chiral edges intact and not connected through the bulk of the sample. Therefore experiments probing these edges still observe the same quantized behavior as if the extra pinned quasi-particles were not there.

\noindent  Therefore, we see that the robustness of the Hall plateau requires several key ingredients beyond the mere existence of a parent ideal incompressible fluid, and a particularly crucial ingredient is that the disorder potential is effective at pinning the charged quasiparticles added to the parent ideal fluid. As we have mentioned, in the traditional Laughlin and Halperin states in standard Landau levels, all charged quasiparticles are drifting particles (such as the Laughlin anyon or the electon) whereas the itinerant particles are all neutral particles (such as the magneto-roton or the Bogoliubov-composite-fermion of the Moore-Read state). Because charged particles are non-itinerant they can be easily pinned by the disorder potential, and Hall plateaus tend to be robust.

However, in the current context of time-reversal invariant pairs of Landau levels with opposite magnetic fields, charged quasiparticles can be itinerant (such as the charged $f_e,f_h$ composite fermions of PH Halperin states discussed in the previous section). Therefore, when a net excess charge is added to the ideal incompressible PH Halperin state so that the filling deviates slightly from the ideal total filling $\nu=1$, the system might be doped with a finite density of itinerant charged quasi-particles which cannot be efficiently pinned by the disorder potential. In the special case of the $f_e,f_h$ particles these are moreover charged fermions with can naturally form a fermi-fluid-like metallic state at finite density. These intinerant charged particles will easily move across the sample and electrically connect the initially disconnected edges through the bulk. The disorder potential will scatter them but it will be much less efficient at pinning and localizing them, as compared to the drifting quasi-particles. This will lead to smooth deviations of the Hall and longitudinal conductivities away from those of the parent ideal PH Halperin state and in particular degrade the quantization of its edge conductance. This is consistent with the experimental observations of Ref.\cite{kang2024observation}, which observed as smooth variation of the longitudinal and Hall resistances away from the pressumed ideal filling of $\nu=-3$ and  no clear sign of a robust Hall plateau. 

Notice that this mechanism might not be as efficient at destroying the Hall plateu of an simple Ising Chern number in which particles occupy a single valley Chern band, such as the state believed to be realized at $\nu=-1$ in Ref.\cite{kang2024observation}. This is because in this case we expect that the simplest itinerant charged quasiparticle is the cooper-pair-like bound state of two electrons in separate valleys.  Notice that such Cooper would have to be made from two electrons on the two valleys, and since one of the valley-resolved Chern bands would be fully occupied in the Ising Chern magnet at $\nu=-1$, then this Cooper-pair would have put an electron on one of the empty and gapped moire bands that is not related by time reversal to the occupied band. Thus such Cooper-pair-like bound states might have a larger gap that single electrons and holes (which are not itinerant), or perhaps no good cooper-pair like bound state of these electrons which is well separated from the continuum might appear at low energies. Thus it might not be energetically favorable to add it to the Ising Chern magnet vacuum when the charge density deviates away from the ideal filling but instead to add usual electrons or holes which are drifting (not itinerant) and thus more easily pinned by the disorder. This is consistent with the observation of the Hall plateau for the state at $\nu=-1$ in Ref.\cite{kang2024observation}. 

In the supplementary material we discuss a toy model for bound states of two particles in the case of opposite and also the same magnetic fields. This model illustrates than that when the bound state is a charged but itinerant particle, there is a competition between the interaction that tries to localize the relative distance between the two charges that experience opposite magnetic fields while delocalizing the average position of the center of the bound state. Thus when the interaction scale associated with binding energy of the two components is larger than the disorder potential that tries to localize its average position, the disorder becomes inefficient at pinning and ultimately localizing the itinerant particle. This picture suggests that disorder is less efficient at localizing bound states of particles whose constituents reside in bands with opposite Chern numbers, even in comparison to completely trivial bands that have no Berry phase geometry.

\textit{\textcolor{blue}{Summary, discussion and outlook}}.
We have demonstrated that a subset of Halperin states where particles are added to an empty band with Chern number $C=1$ and holes are added to a filled band with opposite Chern number $C=-1$, have exactly zero Hall conductivity in spite of spontaneously breaking time-reversal symmetry. These states are those for which the intra-flavor ($m$) and inter-flavor ($n$) exponents are related as $m=n+2$. Moreover, in the presence of separate particle number conservation for the two flavors, they feature a fractional helical edge conductance of $e^2/2h$ per edge. Therefore these states behave from the charge transport point of view as half of the standard quantum spin Hall states \cite{kane2005quantum,kane2005z,bernevig2006quantum}, and are, therefore, consistent with those observed in moir\'e tMoTe$_2$ in Ref.\cite{kang2024observation}.

We have also emphasized, using these PH-Halperin states as examples, a crucial difference between the standard setting of multi-component systems in Landau levels with common magnetic fields, and our setting of pairs of Landau levels with opposite magnetic fields. Namely, in standard Landau levels all itinerant quasi-particles are neutral (e.g. excitons in quantum Hall ferromagnets, magneto-roton of Laughlin state, Bogoliubov composite-fermion of Moore-Read state), and charged particles are drifting particles which experience Lorentz force. However, in the current setting of pairs of Landau levels with opposite magnetic fields it is possible to have charged particles that are itinerant when their charge is equally split between the valleys experiencing effectively opposite magnetic fields. In particular, we have seen that the Bogoliubov composite fermion of the 331 PH-Halperin state is a charged itinerant spinless fermion in the current setting. The disorder potential can scatter these particles, but it is much less efficient at pinning and localizing them, because they are itinerant. As a result one expects that if these particles are added to the parent ideal PH-Halperin state for example by changing the electron density slightly away from the precise filling, the disorder potential is less efficient at localizing them and they can make a conducting metallic fluid that will degrade the quantization of the conductance. This is consistent with the experimental observation of no robust Hall plateau surrounding the ideal filling fraction where the fractional quantum spin Hall effect is observed, but instead a smooth variation of the Hall resistance away from zero as function of filling in Ref.\cite{kang2024observation}.

There are several important open questions for future studies which we would like to mention. One set of questions pertains to investigating experimental probes which could distinguish the PH-Halperin states from other possible states. In this regard, we would like to highlight that the PH-Halperin states we have discussed have quasi-particles with minimal charge equal to $1/2$ (see Fig.\ref{fig2}). This is in contrast with typical paired states realized in half-filled Landau levels (including the standard Halperin 331 and the Moore-Read states) which have quasiparticles with minimal charge $1/4$. For example, measuring this charge would distinguish our state from a state in the same universality class of a time-reversal invariant pair of Moore-Read states constructed in the $C=\pm 1$ bands. Measuring the fractional charge of quasi-particles is difficult but there are important precedents from current-noise experiments in quantum Hall settings \cite{saminadayar1997observation,de1998direct}. Anyon interferometers could also be used to measure the quasi-particle charge and statistics \cite{nakamura2020direct}.
Other interesting observables to further consider include the spin/valley Hall-drag, which is expected to be quantized according to Eqs.\eqref{Hallmatrix}, and could be probed by inter-facing the PH-Halperin state with another magnetic state that would act as spin battery inducing a non-zero spin-Hall-voltage. Other possibilities include also measuring the thermal Hall conductance \cite{banerjee2017observed,banerjee2018observation} which is expected to ideally be non-zero and the same as an integer quantum Hall effect, since the K-matrix from Eq.\eqref{Kmatrix} has two negative and one positive eigen-value. This would distinguish the PH-Halperin states from those with time-reversal symmetry, which are ideally expected to have vanishing thermal Hall conductance.

Another important open problem is to understand which microscopic models and interactions stabilize the PH-Halperin states in tMoTe$_2$. Clearly there are many non-trivial realistic aspects of moir\'e tMoTe$_2$ that our current discussion is missing. However we would like to highlight that even understanding the competition of the PH-Halperin states and other states in ideal pairs of Landau levels with opposite magnetic fields remains relatively unexplored. We would like to comment in this regard that Ref.\cite{stefanidis2020excitonic} demonstrated that for simple toy models of short-ranged gaussian interactions with a range comparable to the moir\'e unit cell, the Ising Chern magnet becomes unstable when the intra-flavor repulsions $V_{\uparrow,\uparrow}$ become about $30\%$ stronger than inter-flavor repulsion $V_{\uparrow,\downarrow}$. References \cite{stefanidis2020excitonic,kwan2022excitonic} proposed that one possible set of states emerging in this setting might be Laughlin states of excitons. But it would be important to examine these possibilities in detailed many-body numerical studies and search for PH-Halperin and other possible states.

Our PH-Halperin states clearly illustrate that for a pair of Chern bands with valley Chern numbers $C=\pm1$, having a gapped state with zero Hall conductivity does not imply absence of valley polarization or time-reversal symmetry. Conversely having zero valley polarization in this same setting does not imply a zero Hall conductivity, as exemplified by the excitonic Laughlin states constructed in Refs.\cite{stefanidis2020excitonic,kwan2022excitonic}, which can be valley unpolarized but have a quantized integer Hall conductivity $\sigma_{xy}=e^2/h$. These examples highlight that in the setting of pairs of valleys with opposite Chern numbers it is important to exercise some caution as the observation of integer quantized Hall conductivities could be secretly disguising non-trivial fractionalized states.

Finally, we would like to close by mentioning that there are other interesting PH-Halperin states that we have not focused on because they do not fit the experimental observations of Ref.\cite{kang2024observation}, but which might appear in other experimental settings and are also interesting theoretically. For example from Eq.\eqref{sigmaxy} we see that another interesting subset of PH Halperin states with integer Hall conductivity, $\sigma_{xy}=e^2/h$, are those with $m=n+1$, such as the PH 332 state, which has fractional valley fillings $\nu_e^{\uparrow}=1/5$ and $\nu_e^{\downarrow}=4/5$. Another interesting states are those with $m=n-1$, such as PH 112, which would have Hall conductivity $\sigma_{xy}=-3e^2/h$, and valley fillings $\nu_e^{\uparrow}=1/3$ and $\nu_e^{\downarrow}=2/3$, and those with $m=n-2$, such as PH 113, which would have Hall conductivity $\sigma_{xy}=-2e^2/h$, and valley fillings $\nu_e^{\uparrow}=1/4$ and $\nu_e^{\downarrow}=3/4$. These states further highlight the aforementioned point that integer quantization of the Hall conductivity might hide behind more interesting states in disguise.

\textit{\textcolor{blue}{Acknowledgements}}. I am especially indebted to my colleague Bernd Rosenow for many important and stimulating discussions on this subject, and, in particular, for teaching me about various aspects of the physics of quantum Hall edges which were crucial for me to gain deep appreciation of the experimental findings of Ref.\cite{kang2024observation}. I am also thankful to the organizers and participants of three conferences which were crucial for gaining early insights for developing these ideas: the ``Workshop
Fractional Quantum Anomalous Hall Effect and Fractional Chern Insulators" at the Max-Planck-Institute for the Physics of Complex Systems in Dresden, the ``FTPI March Meeting" at the Physics Department of the University of Minnesota supported by the Simons Foundation, and the 2024 March Meeting in Minneapolis of the American Physical Society. In these conferences I enjoyed stimulating discussions in particular with Yang Zhang, Ady Stern, Liang Fu, Allan MacDonald, Cenke Xu, T. Senthil, and Andrei Bernevig. I am also thankful for the support from the Deutsche Forschungsgemeinschaft (DFG) through research grant project number 518372354.

\bibliography{references}

\clearpage
\widetext

\begin{center}
\textbf{\large Supplementary Material: Halperin States of Particles and Holes in Time Reversal Invariant Pairs of Landau Levels and Their Possible Presence in Moir\'e MoTe$_2$}
\end{center}

\setcounter{equation}{0}
\setcounter{figure}{0}
\setcounter{table}{0}
\setcounter{page}{1}
\makeatletter
\renewcommand{\theequation}{S\arabic{equation}}
\renewcommand{\thefigure}{S\arabic{figure}}

\newcounter{proplabel}


\section{A toy model of competition of interactions and localization of charged itinerant particles in pairs of time-reversal invariant Landau levels.}

\noindent  To illustrate this mechanism more concretely, let us consider a toy-model of two particles with different charges $q_1,q_2$ and experiencing different magnetic fields $B_1,B_2$. The operator that generates translations within a Landau level is the magnetic momentum, which for each of these particles takes the form:

\begin{equation} \label{eq:PHHalp}
 {\bf Q}_{a}^1=\pmb{\pi}_{a}^1+B_1 q_1 \epsilon_{a b}{\bf r}_{b}^1 , \  {\bf Q}_{a}^2=\pmb{\pi}_{a}^2+B_2 q_2 \epsilon_{a b}{\bf r}_{b}^2.
\end{equation}

\noindent These particles have non-commutative magnetic momenta:

\begin{equation} \label{eq:PHHalp}
[ {\bf Q}_{a}^i, {\bf Q}_{b}^j]= i  \hbar \epsilon _{a b} \delta^{ij} q_i B_i .
\end{equation}

\noindent Now the operator that generates the translations of these two particles together while preserving their relative positions is the total magnetic momentum, which has commutation relations:

\begin{equation} \label{eq:PHHalp}
{\bf Q}_{a}\equiv{\bf Q}_{a}^1+{\bf Q}_{a}^2, [{\bf Q}_{a}, {\bf Q}_{b}]= i  \hbar \epsilon _{a b} \delta^{ij} (q_1 B_1+q_2 B_2).
\end{equation}

\noindent From the above it follows that a particle made as a bound state of these two particles would be itinerant if the above momenta commute and otherwise it would be drifting. For the standard Landau level setting where each particle experiences equal magnetic fields $B_1=B_2$, we see that only charge-neutral bound states ($q_1=-q_2$) are itinerant while charged bound states are drifting. On the other hand for the case of pairs of time-reversal invariant Landau levels with opposite fields $B_1=-B_2$, we have the converse situation where charge neutral ($q_1=-q_2$) bound states are drifting, whereas a charged bound state with equal charges on the two components ($q_1=q_2$) is itinerant.

\noindent Despite having a commutative total momentum, these itinerant bound states in time-reversal-invariant pairs of Landau levels have non-trivial differences with respect to trivial intinerant particles and their bound states in a trivial band without any Berry phases whatsoever. To see this, let us consider introduce the guiding center position (or projected position) operators for each particle defined as follows:

\begin{equation} \label{eq:PHHalp}
{\bf R}_{a}^i=-\frac{\epsilon _{a b}}{q_iB_i} {\bf Q}_{b}^i={\bf r}_{a}^i-\frac{\epsilon _{a b}}{q_i B_i} \pmb{\pi}_{b}^i.
\end{equation}
 
\noindent We can then define a relative distance ${\bf d}$ and average position ${\bf R}$ of the pair of particles as follows:
 
\begin{equation} \label{eq:PHHalp}
{\bf d} \equiv {\bf R}^1-{\bf R}^2,  {\bf R} \equiv \frac{{\bf R}^1+{\bf R}^2}{2},
\end{equation}

\noindent These operators obey the following algebra:

\begin{equation} \label{eq:PHHalp}
\left[ {\bf d}_a,{\bf d}_b\right]=  \left(\frac{1}{q_1B_1}+\frac{1}{q_2B_2} \right) i \hbar \epsilon _{a b},
\end{equation}

\begin{equation} \label{eq:PHHalp}
\left[ {\bf R}_{a},{\bf R}_{b}\right]=\frac{1}{4} \left(\frac{1}{q_1B_1}+\frac{1}{q_2B_2} \right) i \hbar \epsilon _{a b},
\end{equation}

\begin{equation} \label{eq:PHHalp}
\left[{\bf R}_{a},{\bf d}_b\right]=\frac{1}{2} \left(\frac{1}{q_1 B_1}-\frac{1}{q_2 B_2}\right) i \hbar \epsilon _{a b},
\end{equation}

\noindent We see therefore that in Landau levels with $B_1=B_2$ the algebras of relative distance and average position commute for bound states of particles with equal charge $q_1=q_2$. However, for time-reversal pairs of Landau levels with $B_1=-B_2$ and bound states of particles with equal charge $q_1=q_2$, the relative distance operator does not commute with the average position operator. The interaction between the particles depends on their relative distance and thus would like to fix this operator, whereas the disorder potential wants to localize the particles and pin their average positions. But the Heisenberg uncertainty prevents optimizing both of these properties simultaneously. There is, therefore, a competition between disorder and interactions. If the interaction is stronger, the system will prefer making more certain the relative distance ${\bf d}$ leading to strong quantum fluctuations on the average position, ${\bf R}$, thus reducing the ability of disorder to localize and pin these particles. This is a delocalization mechanism that would not be present in trivial bands with fully trivial Berry phases where projected position operators commute.

\noindent To illustrate this let us consider a toy harmonic Hamiltonian of the form:

\begin{equation} \label{Harmonic}
H=\sum_a \frac{U}{l^2_B} {\bf d}_{a} {\bf d}_{a}+\frac{V}{l^2_B} \left({\bf R}_{a}^1 {\bf R}_{a}^1+{\bf R}_{a}^2 {\bf R}_{a}^2\right)
\end{equation}

\noindent the term proportional to $U$ is a cartoon for an attractive interaction potential that tries to minimize the distance between the particles and $V$ is a cartoon for a disorder potential trying to pin the particles at the origin. Let us assume that the two particles have the same charge $q_1=q_2=q$, but let us contrast two cases: (a) when $B_1=B_2=B$ and (b) $B_1=-B_2=B$. We have taken a unit length $l_B=(\hbar/|qB|)^{1/2}$. In case (a) one finds two decoupled Harmonic oscillators one for the ${\bf d}$ and one ${\bf R}$ variables with frequencies and mean fluctuations in the ground state given by:

\begin{equation} \label{Harmonic}
\hbar \omega_{\bf d}=4 U +V, \left\langle {\bf d}_x^2+{\bf d}_y^2\right\rangle=2 l_B^2,
\end{equation}

\begin{equation} \label{Harmonic}
\hbar \omega_{\bf R}=V, \left\langle {\bf R}_x^2+{\bf R}_y^2\right\rangle=\frac{l_B^2}{2},
\end{equation}

\noindent We see therefore that there is no competition between interactions and disorder in this case, and the relative and average position can be simultaneously localized on a similar length scale comparable to the magnetic length $l_B$ irrespective of the values of $U$ and $V$. Now in case (b) we find two degenerate oscillators, one generated by the pair of operators ${\bf d}_x,{\bf R}_y$, and another generated by the pair ${\bf d}_y,{\bf R}_x$. The frequency of these oscillators and fluctuations in the ground state are:

\begin{equation} \label{Harmonic}
\hbar \omega=\sqrt{V \left(4 U+V\right)}, \end{equation}

\begin{equation} \label{Harmonic}
\left\langle {\bf R}_x^2+{\bf R}_y^2\right\rangle=\frac{l_B^2}{2} \sqrt{\frac{4 U+V}{V}},  \left\langle {\bf d}_x^2+{\bf d}_y^2\right\rangle=2 l_B^2 \sqrt{\frac{V}{4 U+V}}.
\end{equation}

\noindent The above clearly illustrates the competing tendencies of interactions and disorder in case (b). When the interaction dominates the disorder ($U\gg V$) the system optimizes the localization of the relative distance ${\bf d}$ at the expense of delocalizing the average position ${\bf R}$.

\end{document}